# Magnetic behavior of cubic $Dy_4RhAl$ with respect to isostructural $Dy_4PtAl$, revealing a novel 4f d-band interaction

K. K. Iyer,[1,2,3,*] S. Matteppanavar,[2,3] S. Dodamani,[2] K. Maiti[1] and E.V. Sampathkumaran[4,#]

[1]Tata Institute of Fundamental Research, Homi Bhabha Road, Colaba, Mumbai – 400005
[2]Dr. Prabhakar Kore Basic Science Research Centre, KLE Academy of Higher Education and Research, Belagavi- 590010
[3]KLE Society's, Basavaprabhu Kore college of Arts, Science & Commerc, Chikodi-591110
[4]Homi Bhabha Centre for Science Education, Tata Institute of Fundamental Research, V.N. Purav Marg, Mankhurd, Mumbai 400088.

## ABSTRACT
We have investigated for the first time the magnetic behaviour of an intermetallic compound, $Dy_4RhAl$, crystallizing in $Gd_4RhIn$-type cubic structure containing 3 sites for rare-earth (R), by several bulk measurements down to 1.8 K. This work is motivated by the fact that the isostructural Dy compound in the $R_4PtAl$ family surprisingly orders ferromagnetically unlike other members of this series, which order antiferromagnetically. The results reveal that the title compound undergoes antiferromagnetic order at about 18 K, similar to other heavy R members of $R_4RhAl$ family, unlike its Pt counterpart, indicating a subtle difference in the role of conduction electrons to decide magnetism of these compounds. Besides, spin-glass features coexisting with antiferromagnetic order could be observed, which could mean cluster antiferromagnetism. The electrical resistivity and magnetoresistance behaviours in the magnetically ordered state are typical of magnetic materials exhibiting antiferromagnetic gap. Features attributable to spin-reorientation as a function of temperature and magnetic field can be seen in the magnetization data.

*iyerkk@gmail.com
#Corresponding author: sampathev@gmail.com

## 1. Introduction

The ternary compounds of the type $R_4TX$ (where $R$ = rare-earths, $T$ = transition metals and $X$ = p-block metals) forming in the cubic $Gd_4RhIn$-type structure [1-4] are of special interest, as these compounds provide an opportunity to understand magnetism due to the interactions between 3 independent crystallographic sites of the rare-earth. We have carried out extensive bulk measurements on many heavy rare-earth members of the families, $R_4PtAl$ (R= Gd, Tb, Dy, Ho and Er) and $R_4RhAl$ (R= Gd, Tb, Ho and Er) and we have observed re-entrant spin-glass behaviour in all the compounds, apart from many other interesting magnetic, thermal, transport and magnetocaloric features [5-10]. A novel observation pertinent to the aim of this article is that the onset of magnetic order is of an antiferromagnetic (AF) type for R= Gd [5], Tb [6], Ho and Er [10], within the $R_4PtAl$ family, whereas Dy member [7] enters into a ferromagnetic (F) state from the paramagnetic state. It is puzzling why $Dy_4PtAl$ behaves differently in this respect. It is therefore of great interest to understand the magnetism of Dy counterpart in another family, viz., $R_4PtAl$, as Gd, Tb, Ho and Er members of this family also order antiferromagnetically with pronounced spin-glass features. With this primary motivation, we have carried out ac and dc magnetization ($M$), heat-capacity ($C$), electrical resistivity ($\rho$), and magnetoresistance (MR) measurements on $Dy_4RhAl$, the results of which are presented in this article. It may be remarked that, subsequent to initial synthesis report of this compound [2] about a decade ago, to our knowledge, there is no further literature on the magnetism of this compound. The point of main emphasis, apart from other findings, is that this compound also orders antiferromagnetically similar to other members in this family - unlike $Dy_4PtAl$ - forming clusters behaving like spin-glass.

## 2. Experimental details

The sample in the polycrystalline form was prepared by melting together stoichiometric amounts of high purity Dy (>99.9%), Rh (99.99%) and Al (>99.99%) in an atmosphere of high purity argon in an arc furnace. Powder x-ray diffraction pattern (XRD) (Cu K$_\alpha$) confirmed single phase of the specimen within the detection limit of this technique (<2%). The results of Rietveld refinement to the cubic phase with the space group $F\bar{4}3m$ are shown in Fig. 1. The lattice constant is obtained to be 13.403(3) Å, which is in good agreement with the reported value of 13.468 Å [2]. We have also characterized the sample by scanning electron microscope (SEM) and we could not detect any additional phase. Energy Dispersive X-ray (EDX) analysis confirmed that the chemical composition is uniform (4:1:1) in the sample. Therefore, within the detection limit of these techniques (<1%), the specimen is a single phase. Dc magnetic susceptibility ($\chi$) (1.8 to 300 K), and isothermal $M$ were performed using a commercial PPMS (Physical Properties Measurement system, Quantum Design) -VSM and ac $\chi$ (with an ac field of 1 Oe) measurements were carried out with the help of a commercial (Quantum Design) superconducting quantum interference device. Heat-capacity and electrical resistivity as well as MR measurements as a function of temperature ($T$) and magnetic field ($H$) were caried out using a commercial PPMS. The measurements in general were carried out for the zero-field-cooled (ZFC) condition of the specimens.

## 3. Results and discussions

In Fig. 2a, we show $\chi(T)$ measured in 100 Oe for the ZFC condition as well as for the field-cooled (FC) condition of the specimen in the low temperature range of interest. There is a distinct peak at $T_N$=18 K suggesting the onset of AF order. In addition to this peak, there are additional weak peaks at about 4 and 12 K, indicating further magnetic anomalies with a lowering of temperature. The 12K-feature is smeared when measured in a field of 5 kOe (Fig. 2b), implying subtle nature of such low-temperature magnetic state, but we do not know whether possible traces of other magnetic impurities below the detection limit of XRD, SEM and EDX cause this feature. Inverse $\chi$ plot for 5 kOe data in Fig. 2c is linear over a wide $T$ range (that is, above 25 K) and a Curie-Weiss fit yielded a value of the effective moment of 11 $\mu_B$/Dy, which is marginally higher than the theoretical value of 10.48 $\mu_B$ for $Dy^{3+}$, usually attributed to a contribution from conduction electron polarization. The paramagnetic Curie temperature is obtained to be -16 K; this magnitude and the sign are in good agreement with AF ordering setting in around 18 K. Finally, the low-field curves for ZFC and FC conditions separate at $T_N$, and the FC curve exhibits an upturn with decreasing temperature typical of cluster spin-glasses [11-13]. All these results already indicate possible antiferromagnetic cluster glass behaviour. We would also like to mention that the χ curves of 100 Oe and 5 kOe do not overlap at $T_N$, but well above $T_N$ only, as shown in the inset of Fig. 2c, and this signals the existence of short-range magnetic correlations before long-range magnetic ordering sets in, as demonstrated for many heavy rare-earth systems [see the articles cited in [9]].

Fig. 3a shows isothermal magnetization at 2, 5 and 10 K. *M* increases gradually with *H* at these temperatures, tending to saturate beyond about 60 kOe. A careful look at the curve suggests that there is an upward curvature around 20-40 kOe and a weak hysteresis is clearly visible as the temperature is lowered to 2 K. This implies that there could be a disorder-broadened first-order field-induced magnetic transition. All these features are consistent with the conclusion that the zero-field state is not ferromagnetic. The values of the magnetic moment at very high fields (as well as the zero-field value following linear extrapolation of high-field linear region) are far less than that of the free ion and this can arise from crystal-field-effects.

In Fig. 4a-b, we show $C(T)$ and $C(T)/T$ behaviour in zero field as well as in the presence of 10, 30 and 50 kOe. In the plot of $C(T)$ versus *T*, in zero-field, there is an upturn below 20 K, followed by a peak at 18 K and a gradual fall thereafter with decreasing temperature. There is no well-defined peak well-below 18 K suggesting the absence of any additional long-range magnetic ordering, but not inconsistent with possible presence of spin-glass freezing. A careful look at the $C/T$ curves reveals a change of slope around 5-10 K in the zero-field curve, thereby implying that there is actually a subtle faster decrease (compared to higher temperature linear region) in heat-capacity attributable to the spin-glass component. The support for the onset of antiferromagnetic order can be seen from the gradual suppression of the peak with increasing *H*. We have also obtained isothermal entropy change, $\Delta S = S(H) - S(0)$, as a function of temperature by integrating the $C/T$ versus *T* curves. The plots of $-\Delta S$ (Fig. 4c) thus obtained for *H*= 30 and 50 kOe exhibit a positive peak at a temperature (about 25 K) slightly above $T_N$; this finding, along with the positive sign, implies [14] that short-range magnetic correlations (inferred from the inset of Fig. 2c) setting in before long-range magnetic order is of a ferromagnetic-type. There is also a weaker peak in the negative quadrant around 18 K, and the sign is consistent with AF order. There are additional sign-crossovers at lower temperatures, which may be consistent with subtle magnetic effects, mentioned above. The peak values are not large, as in other Rh compounds [8, 9]. The values for *H*= 10 kOe are too small to attach any significance to the sign changes with *T*. Finally, we are able to fit the $C(T)$ data below 7 K to the functional form, $C = \beta T^3 + \alpha T^n e^{-\Delta/T}$, where α, β and *n* are constants, and Δ is the spin-gap between the lower and upper band of the spin wave spectrum. The value of Δ and *n* are found to be ~2.2 K and ~0.8.

In order to explore possible presence of spin-glass features, we have measured ac susceptibility with four frequencies (υ). We show the real (χ') and imaginary (χ") parts in Fig. 5. Since the curves are noisy for 133 and 1333 Hz, we show the curves for 1.3 and 13 Hz only for 5 kOe. Though the data for higher frequencies are noisy, a υ-dependence of the peak is transparent in the real part in the vicinity of $T_N$ (for instance, ~20 and ~24 K for 1.3 and 1333 Hz respectively), attributable to spin-glass freezing. From the values of the peak temperature ($T_p$) for the two extreme frequencies, the magnitude of the factor, $\Delta T_p / T_p \Delta(\log υ)$, is derived to be about 0.07, which is rather large compared to that for canonical spin-glasses (which is <0.01) [15]. The fact that the $T_p$ at the lowest frequency (1.3 Hz) matches with the value of $T_N$, suggests that spin-glass freezing sets in simultaneously with antiferromagnetic ordering. Combined with the observation that there is an upturn in the low-field FC χ in the magnetically ordered state (mentioned above) with decreasing temperature, we infer that the spin-glass is of a cluster-type, arising from antiferromagnetic clusters. However, there is no such feature in χ" down to 2 K (shown for two frequencies only). The presence of a worthwhile signal in χ" is a necessary criterion for spin-glass freezing [15]. We therefore infer that the spin-glass dynamics of the AF clusters is so weak that it escapes detection by χ" measurement. An application of a dc field (say, 5 kOe) suppresses the peak in χ' marginally (in sharp contrast to the complete suppression in conventional spin-glasses) and this endorses weakness of spin-glass feature. Finally, there is a drop in χ' below 5 K, as in dc χ data, which is also distinctly frequency dependent behaving like the one at 18 K, Thus, down to 2 K, spin-glass behaviour coexists with AF as soon as long-range magnetic ordering sets in.

In order to render further support to spin-glass freezing, time ($t$) dependence of isothermal remnant magnetization ($M_{IRM}$) was measured below $T_N$. For this purpose, the sample was cooled to desired temperature in zero field, and then a field of 5 kOe was switched on for 5 mins. Immediately after switching off the field, $M_{IRM}$ was tracked as a function of time ($t$), and the results are shown in Fig. 3b for 4, 8 and 15 K. It is clear that there is a slow decay of $M_{IRM}$ (essentially logarithmically after waiting for about a minute) at all temperatures, characteristic of many spin-glass systems; however, the magnitude of $M_{IRM}$ at $t=0$ decreases with increasing $T$, reflecting a change in the strength of spin-glass component with respect to antiferromagnetic part. Above $T_N$, say at 25 K, no such decay was observed.

The behaviour of $\rho(T)$ in the presence of various external fields (0, 30, 50, 70 and 100 kOe) and isothermal MR, defined as $[\rho(H)-\rho(0)]/\rho(0)$, at 5, 8, 12, 16 and 25 K are shown in Figs. 6a and 6b respectively. The derivative, $d\rho/dT$, in the paramagnetic state is metallic for all fields. The magnitude of ρ however undergoes pronounced depression with increasing $H$ as the $T$ is lowered, revealing a gradual increase of MR as the magnetic ordering is approached. In the past, several rare-earth intermetallics have been demonstrated to show this behaviour due to a peculiar magnetic precursor effect [16], which is recently understood in terms of an interplay between indirect exchange interaction and magnetic frustration [17]. With a further lowering of temperature, an upturn is observed as soon as magnetic ordering sets in (in zero field). This establishes the formation of magnetic Brillouin-zone formation, consistent with antiferromagnetic nature of the magnetic order [18]. This feature undergoes a gradual suppression with increasing $H$. Naturally, MR is large in the magnetically ordered state. In order to get a better picture of the magnitude of MR, the readers may see the isothermal MR curves, shown in Fig. 6b. A noteworthy finding in this figure is that there is a very broad step around 20-50 kOe at 5 K and the curve is weakly hysteretic supporting the conclusion from $M(H)$ data. This feature gradually gets smeared with increasing $T$. At 25 K, a quadratic field-dependence expected for paramagnets could be seen.

4.  Conclusions

Several bulk measurements establish that the compound $Dy_4RhAl$ undergoes antiferromagnetic order unlike its counterpart in Pt series. The present results therefore offer an indirect support for the fact that the ferromagnetic behaviour of $Dy_4PtAl$ is interesting and unusual. It is possible that the Pt 5d orbital, which is more spatially extended compared to Rh 4d orbital plays a role by strong interaction with crystal-field-split anisotropic orbitals of Dy for the anomalous magnetism of $Dy_4PtAl$. Moreover, 5d electrons are relatively less correlated with stronger spin-orbit coupling than the 4d electrons. This suggests an important role of conduction electrons interacting with the Dy 4f electrons in determining the magnetism of these systems.

We hope that a comparative study of both these Dy compounds by spin-polarized band structure calculations, electron spectroscopy and neutron diffraction (to determine magnetic structure), would offer an ideal opportunity to throw light on the role of 5d band correlations on the magnetism of strictly localized 4f orbital. Inelastic neutron scattering studies are also warranted to learn about crystal-field scheme in these materials.

There are various other interesting observations with respect to the properties of the title compound, viz., evidence for antiferromagnetic energy gap formation, weak spin-glass behaviour of antiferromagnetic clusters, large magnetoresistance not only in the magnetically ordered state but also well above $T_N$ extending to $T>3T_N$, spin-orientation effects as a function of $T$ and $H$. It is of interest to explore theoretically how the interaction among 3 sites of rare-earth plays a role in deciding the magnetic behaviour.

**Acknowledgements**

E.V.S. acknowledges the support of Atomic Energy Department, Government of India, by way of awarding Raja Ramanna Fellowship. K.M. thanks financial support from BRNS, DAE under the DAE-SRC-OI program. SM thanks support from Vision Group on Science and Technology-GRD No 852.

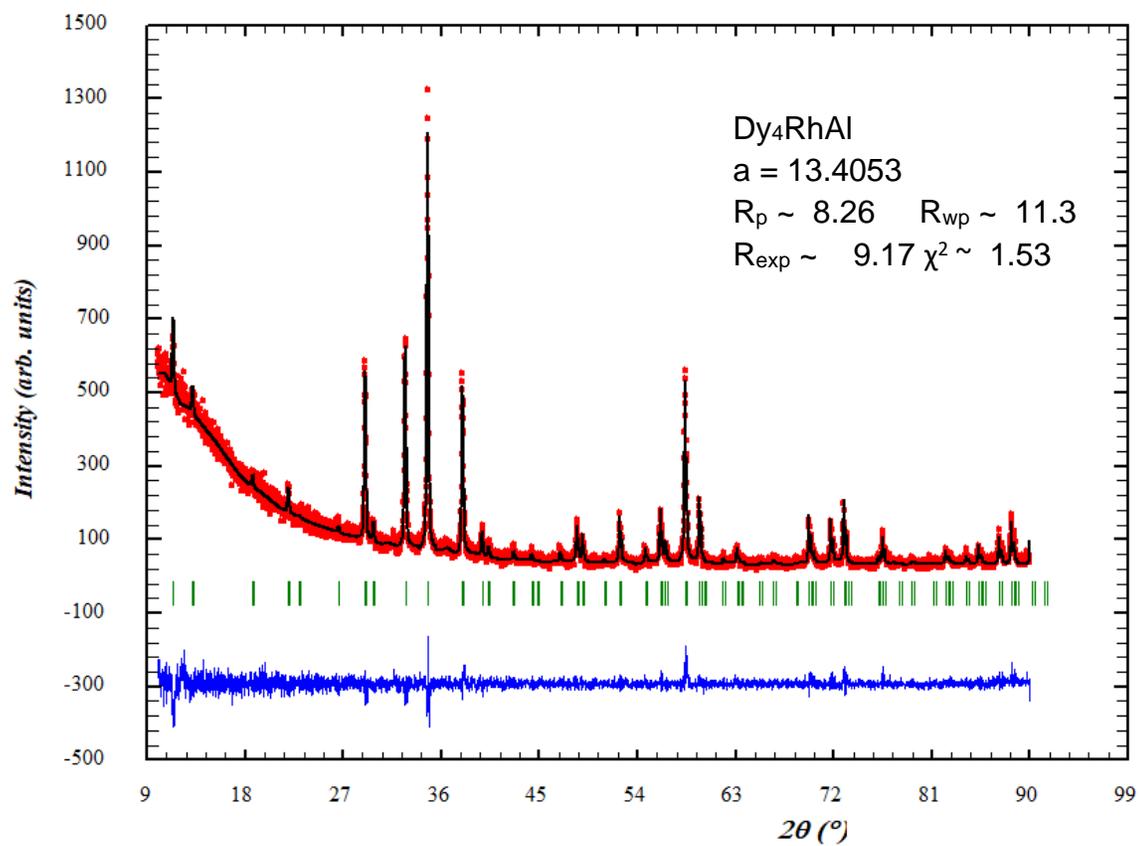

**Fig.1.** X-ray diffraction pattern of Dy$_4$RhAl at room temperature, along with Rietveld fitting results.

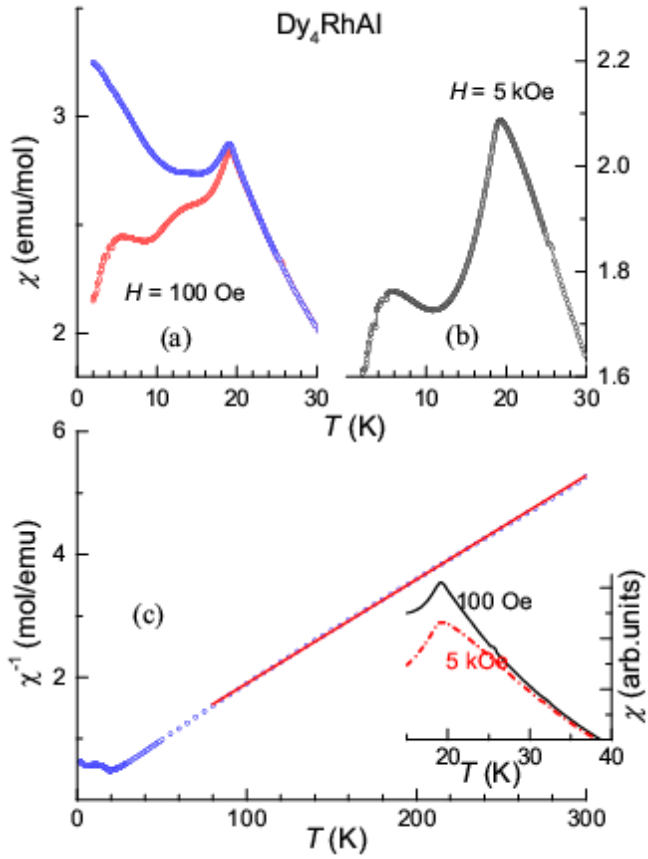

**Fig. 2.** The temperature dependent (a, b) dc magnetic susceptibility measured in 100 Oe (for zero-field-cooled and field-cooled condition) and in 5 kOe (zero-field-cooled) in the low temperature range and (c) inverse susceptibility with the continuous line representing Curie-Weiss fit, for $Dy_4RhAl$. Inset compares the plots of susceptibility measured in 100 Oe and 5 kOe in the vicinity of $T_N$ to bring out the existence of short-range magnetic correlations.

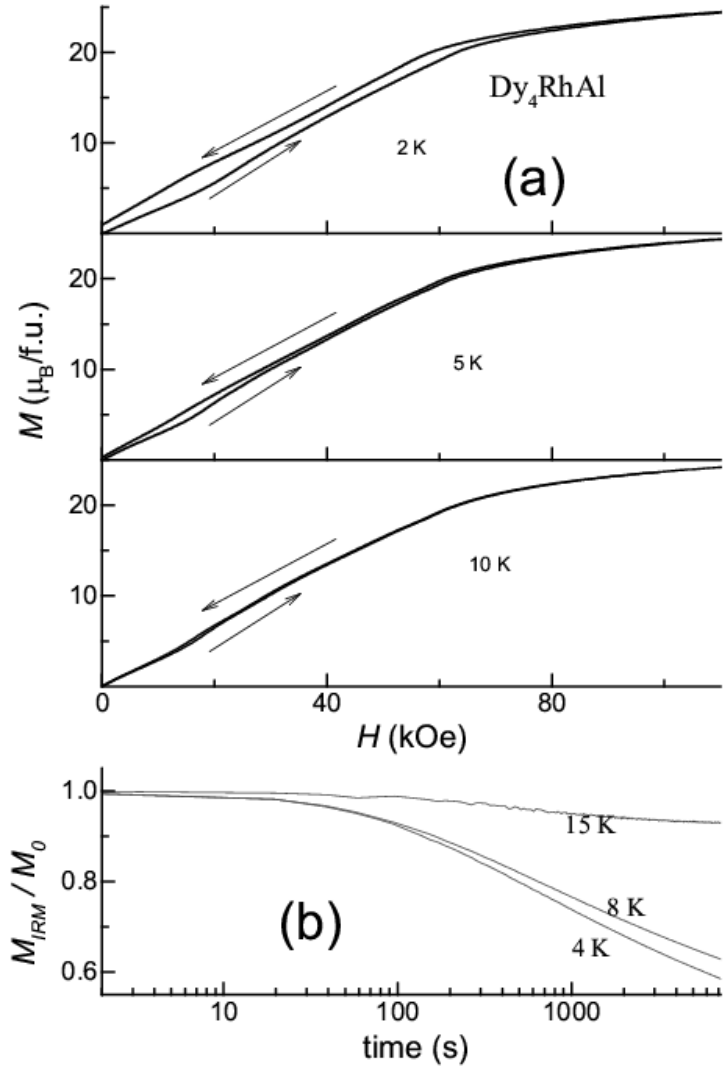

Fig. 3. (a) Isothermal magnetization at 2, 5 and 10 K, and (b) isothermal remanent magnetization as a function of time, normalized to respective values (~ 0.5, 0.25, 0.05emu/g) immediately after the field was switched off, at 4, 8, and 15 K for $Dy_4RhAl$.

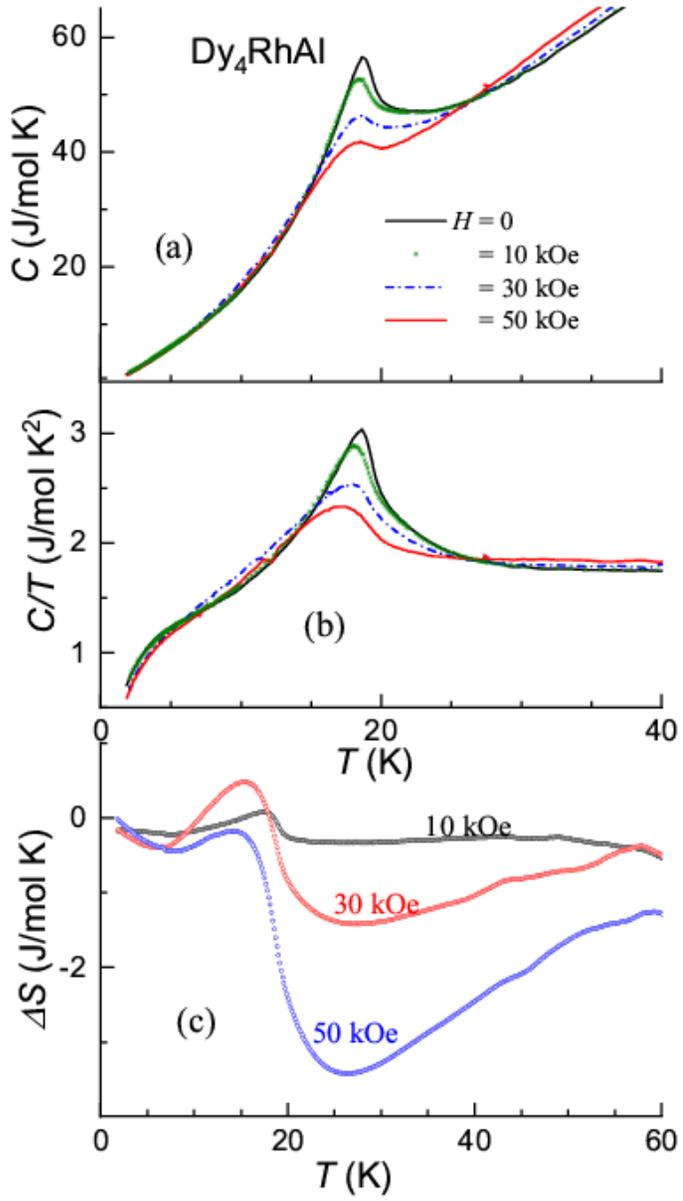

Fig. 4. The plots of (a) heat-capacity ($C$), (b) heat-capacity divided by temperature and (c) isothermal entropy change ($\Delta S$) as a function of temperature (for final fields 10, 30 and 50 kOe) below 40 K for $Dy_4RhAl$.

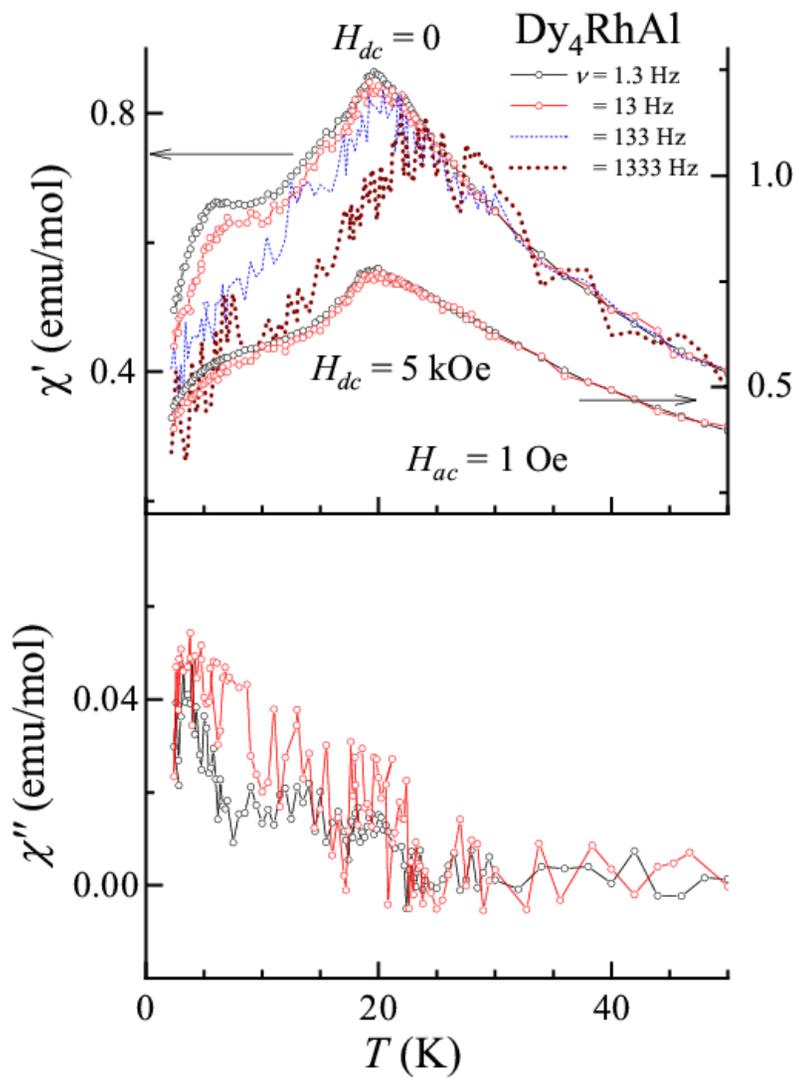

Fig. 5. Real (χ' and imaginary (χ") parts of ac susceptibility for $Dy_4RhAl$. The lines through the data points are guides to the eyes.

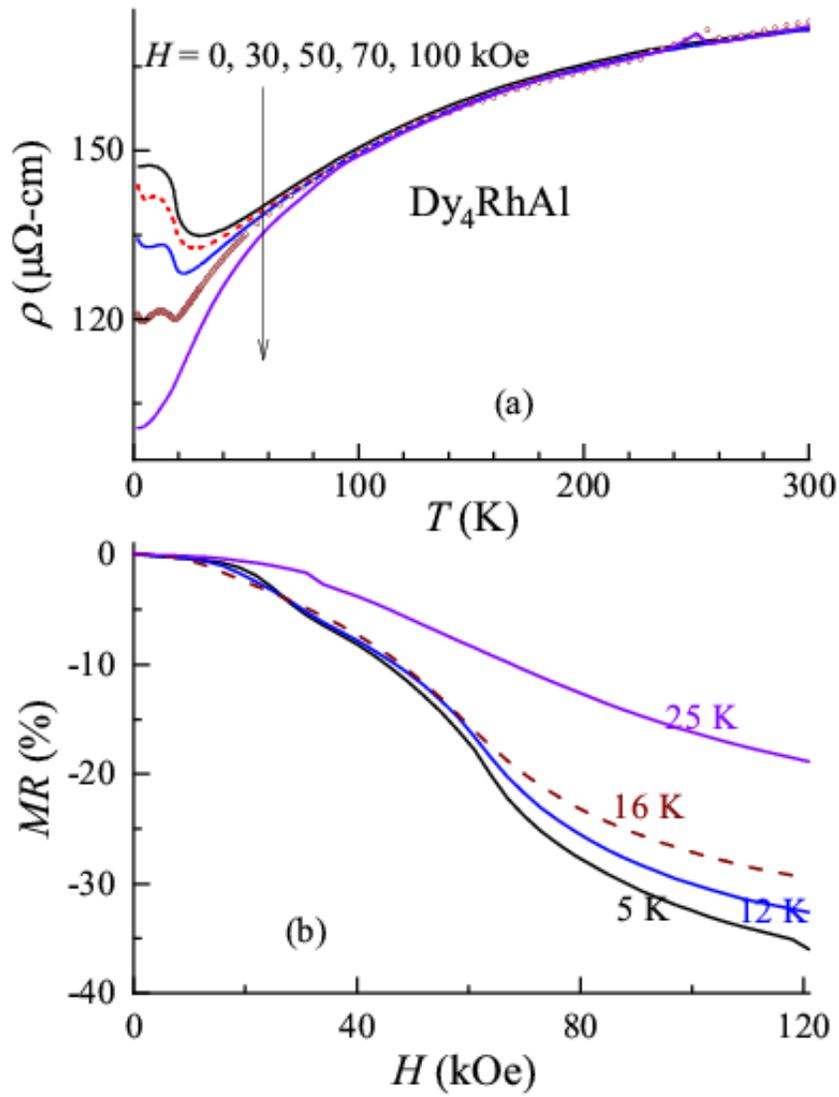

Fig. 6. (a) Electrical resistivity as a function of temperature in several fields, and (b) isothermal magnetoresustance at some temperatures for Dy$_4$RhAl.